\documentclass[%
reprint,
amsmath,amssymb,
aps,
pre,
floatfix,
]{revtex4-2}

\usepackage{hhline}
\usepackage{graphicx}
\usepackage{tabularx}
\usepackage{xcolor}
\usepackage{makecell}
\usepackage{siunitx}
\usepackage{hyperref}

\begin{document}

\title{Load-Dependent Power-Law Exponent in Creep Rupture of Heterogeneous Materials}

\author{Chloé Braux}
 \affiliation{Universite Claude Bernard Lyon 1, CNRS, Institut Lumière Matière, UMR5306, F-69100, Villeurbanne, France}
\author{Antoine Bérut}
 \affiliation{Universite Claude Bernard Lyon 1, CNRS, Institut Lumière Matière, UMR5306, F-69100, Villeurbanne, France}
\author{Loïc Vanel}
 \affiliation{Universite Claude Bernard Lyon 1, CNRS, Institut Lumière Matière, UMR5306, F-69100, Villeurbanne, France}

\begin{abstract}
Creep tests on heterogeneous materials under subcritical loading typically show a power-law decay in strain rate before failure, with the exponent often considered material-dependent but independent of applied stress. By imposing successive small stress relaxations through a displacement feedback loop, we probe creep dynamics and show experimentally that this exponent varies with both applied load and loading direction. Simulations of a  disordered fiber bundle model reproduce this load dependence, demonstrating that such models capture essential features of delayed rupture dynamics.
\end{abstract}


\maketitle


Understanding how heterogeneities influence the mechanical response of materials remains a fundamental challenge, particularly in complex microstructured and amorphous systems. A central open question is the nature of slow rupture dynamics under subcritical loading— i.e., loads below the ultimate tensile strength—that lead to delayed failure over extended timescales. Creep tests on heterogeneous materials reveal that, under constant load, the strain rate typically decreases following a power law $\dot{\varepsilon} \propto t^{-\alpha}$ before accelerating toward failure~\cite{Wyatt1951, Cottrell1952, Nechad2005,Rosti2010, Koivisto2016, Makinen2023, Leocmach2014, Bauland2023, Karobi2016, Pommella2020, Aime2018, Cho2022}, a behavior reminiscent of Andrade creep in metals~\cite{Andrade1910}. The power-law exponent $\alpha$ appears to be material-dependent, typically ranging from 0.4 to 1 for different materials~\cite{Wyatt1951, Cottrell1952, Nechad2005,Rosti2010, Koivisto2016, Leocmach2014, Bauland2023, Karobi2016, Pommella2020, Aime2018, Cho2022}. Some studies suggest a connection between this exponent and that characterizing the frequency-dependent linear viscoelastic response~\cite{Leocmach2014, Bauland2023}, though this relation breaks down beyond the linear regime~\cite{Cho2022}. Nevertheless, $\alpha$ is generally considered stress-independent~\cite{Leocmach2014, Cho2022}, even though some authors report a qualitative decrease when temperature or applied force is increased~\cite{Cottrell1952}. The power-law behavior of the strain rate has been reproduced in Disordered Fiber Bundle Models (DFBMs), driven by viscoplastic flow~\cite{Hidalgo2002, Jagla2011} or thermally activated rupture~\cite{Politi2002, Saichev2005, Fusco2013, Roy2022, Weiss2023} as well as in mesoscopic models of disordered materials~\cite{Moorcroft2013, Miguel2002, Merabia2016, Lockwood2024}. Those models often exhibit an exponent $\alpha$ that varies with temperature and material disorder, but some of them report no dependence on the applied load~\cite{Miguel2002, Merabia2016, Lockwood2024} while others do~\cite{Weiss2023}.

In this article, we experimentally investigate creep in disordered materials and show that the power-law exponent governing strain rate decay depends on both the material and loading direction, emphasizing the impact of anisotropy on creep dynamics. We further find that this exponent varies with applied load. Supporting these results, numerical simulations of a one-dimensional Disordered Fiber Bundle Model reproduce the load dependence, indicating that such models capture essential features of the observed behavior.

\section{Sample characterization and experimental procedure}

\begin{figure}[ht!]
    \centering
     \includegraphics[width=0.9\linewidth]{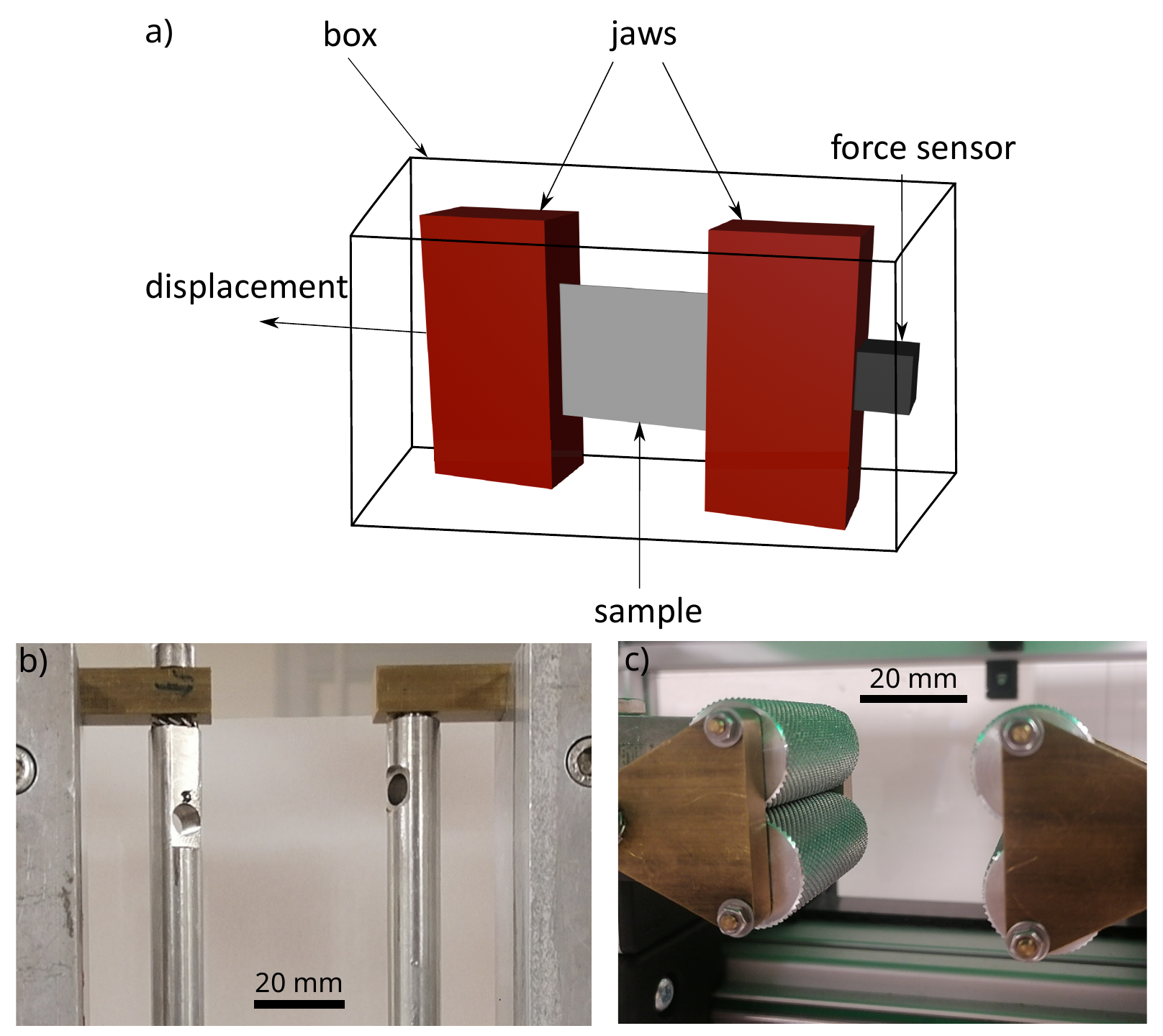}
    \caption{Experimental set-up: custom made tensile apparatus (a). On one side a motor pulls on the sample to impose a deformation $\varepsilon$, while a force sensor measures the resulting load $\sigma$ on the material. The force sensor can be used in a feedback loop to impose a constant load on average. The sample is maintained either by two cylindrical rollers in the case of paper samples (b) or by two self-locking jaws in the case of PDMS samples (c). The entire experimental setup is placed in a semi-hermetic box to control the air humidity.}
    \label{fig:tensile tests set-up}
\end{figure}

\begin{figure}[ht!]
    \centering
     \includegraphics[width=\linewidth]{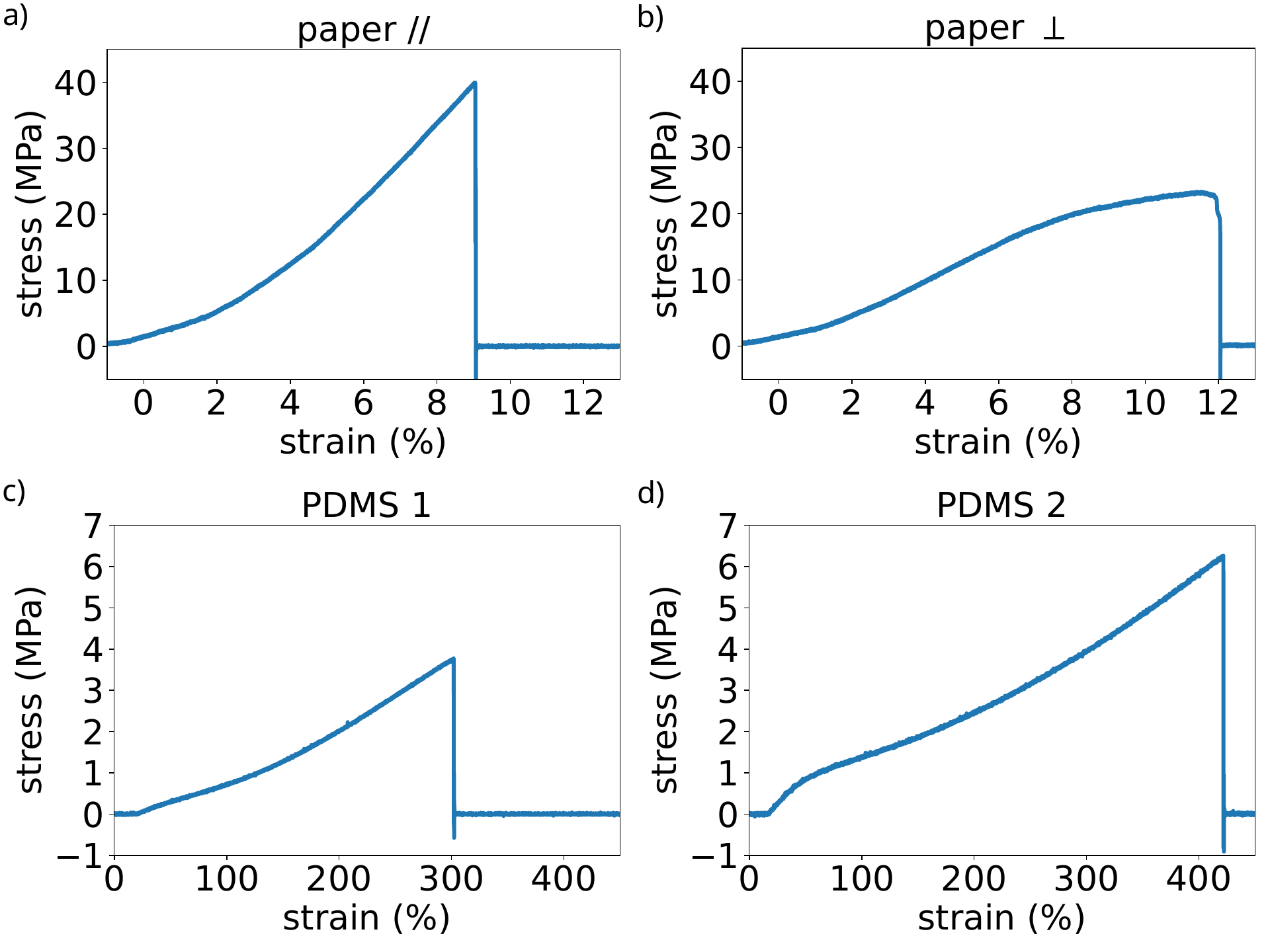}
    \caption{Comparison of the mechanical behavior in tensile test for fax paper in parallel direction (a) and in perpendicular direction (b), and for two kinds silicon elastomers from manufacturers \textit{Gteek} (c) and \textit{GoodFellow Inc} (d).}
    \label{fig:tensile tests curves}
\end{figure}

Our experiments use paper samples (104 mm × 208 mm) cut from fax paper sheets and polydimethylsiloxane (PDMS) samples (100 mm × 10 mm) prepared from two commercial silicone elastomer sheets: 1.4 mm-thick supplied by \textit{GTeek} (``PDMS 1''), and 1 mm-thick supplied by \textit{Goodfellow Inc} (``PDMS 2''). Samples were uniaxially elongated using two custom tensile test apparatus, tailored to each material (see Fig.~\ref{fig:tensile tests set-up}). For paper, the tensile machine uses two rollers as jaws. The paper is aligned on a roller with adhesive tape and then wrapped two turns to secure it by friction. The maximum velocity induced by the motor is \qty{50}{\micro\metre\per\second} and the minimal step is \qty{2.5}{\nano\metre}. The displacement range of this apparatus is limited to about \qty{5}{\centi\metre}, which is enough to break paper samples but makes it unsuitable for PDMS samples, which can support high deformations. The tensile test machine for PDMS has self-locking jaws, a travel range of \qty{1.50}{\metre}, a maximum velocity of \qty{1}{\metre\per\second} and a minimal step of \qty{54.35}{\micro\metre}. Both machines are equipped with a force sensor from \textit{PM instrumentation} (respect. SM \qty{500}{\newton}, and SML \qty{220}{\newton}) enabling a feedback loop on the displacement with a frequency of \qty{25}{\hertz}. For paper samples, we added an hermetic box to better control humidity. All experiments where performed at ambient temperature and $\sim$ \qty{50}{\percent} humidity. Because paper is a fibrous, anisotropic material, its mechanical response depends on the loading direction. Therefore, we divided the samples in two subcategories: paper elongated parallel (paper //) or  perpendicular (paper $\perp$) to the roll direction. Strain-stress curves obtained for different samples are shown in Fig.~\ref{fig:tensile tests curves}. Both PDMS samples exhibit similar behavior during a tensile test. Only the paper in perpendicular configuration exhibits a visible plastic deformation before rupture. Since PDMS is more stretchable than paper, its deformation at break is about 35 times larger ($\varepsilon_{r} \simeq \qtyrange{150}{200}{\percent}$ for PDMS, and $\varepsilon_{r} \simeq \qty{5}{\percent}$ for paper) while its ultimate tensile stress is $\sim 8$ times smaller than that of paper ($\sigma_r \simeq \qtyrange{4}{6}{\mega\pascal}$ for PDMS, and $\sigma_r \simeq \qtyrange{20}{40}{\mega\pascal}$ for paper).

\begin{figure*}[ht!]
    \centering
    \includegraphics[width=0.9\linewidth]{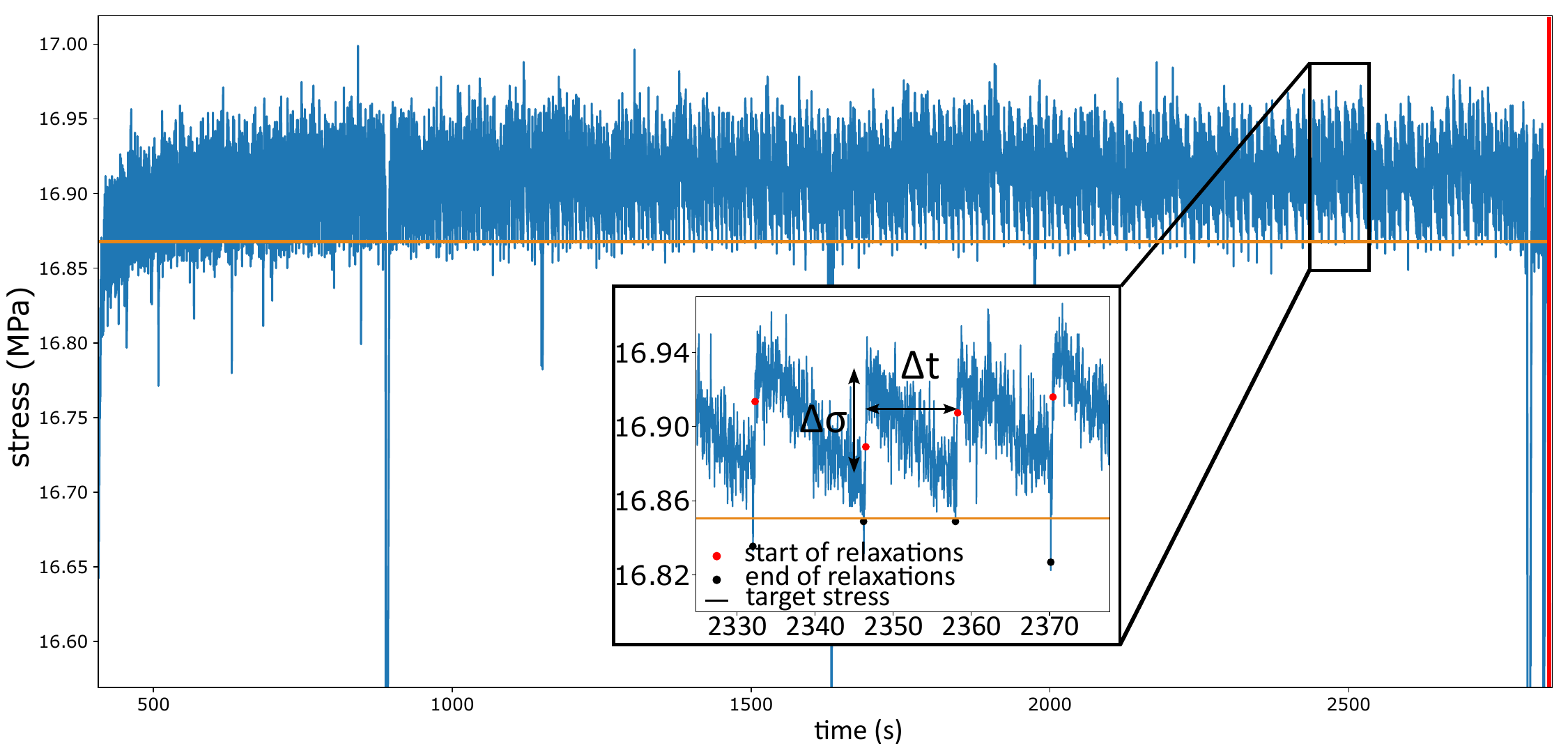}
    \caption{Subcritical rupture process in paper. Here, the sample is loaded with a target stress of \qty{16.87}{\MPa} (orange horizontal line), maintained at the target stress by a feedback mechanism until macroscopic failure at $\tau_{c}$ $\sim$ \qty{3000}{\second} (red vertical line). The inset shows a zoom on a few successive relaxations and the graphical definition of their duration $\Delta t$ and their amplitude $\Delta \sigma$.}
    \label{fig:subcritical process}
\end{figure*}

\section{Dynamics of creep rupture}

We study subcritical rupture by loading a sample to a prescribed force, and maintaining this force constant until failure. An example of the stress signal \footnote{Stresses are computed by dividing the force by the initial section of each sample.} is shown in Fig.~\ref{fig:subcritical process}. The signal has three main parts: initially (not shown on the curve) the sample is loaded to reach the target stress $\sigma_t$ (here \qty{16.87}{\MPa}), taking about \qty{500}{\second}. In the second part, usually the longest one, the stress is maintained near the target using a displacement feedback loop and the sample exhibits creep at constant stress. Finally, the third part of the signal is when the sample breaks (here at time $\tau_c \sim$ \qty{3000}{\second}). To probe the creep dynamics at constant stress, we look more closely at the stress signal: small successive relaxations occur due to the feedback mechanism (see inset of Fig.~\ref{fig:subcritical process}). Indeed, when the stress exceeds the target, the motor stops and the samples's elongation is constant, causing stress relaxation. To counterbalance this, the motor pulls one displacement step $\Delta u$ (here $\Delta u$ = \qty{2.5}{\micro\metre}) as soon as the stress falls below the target. The stress then increases above the target and relaxes again, repeating the process. The successive relaxations can be characterized by their amplitude $\Delta \sigma$ and their duration $\Delta t$.

\begin{figure}[ht!]
    \centering
    \includegraphics[width=0.99\linewidth]{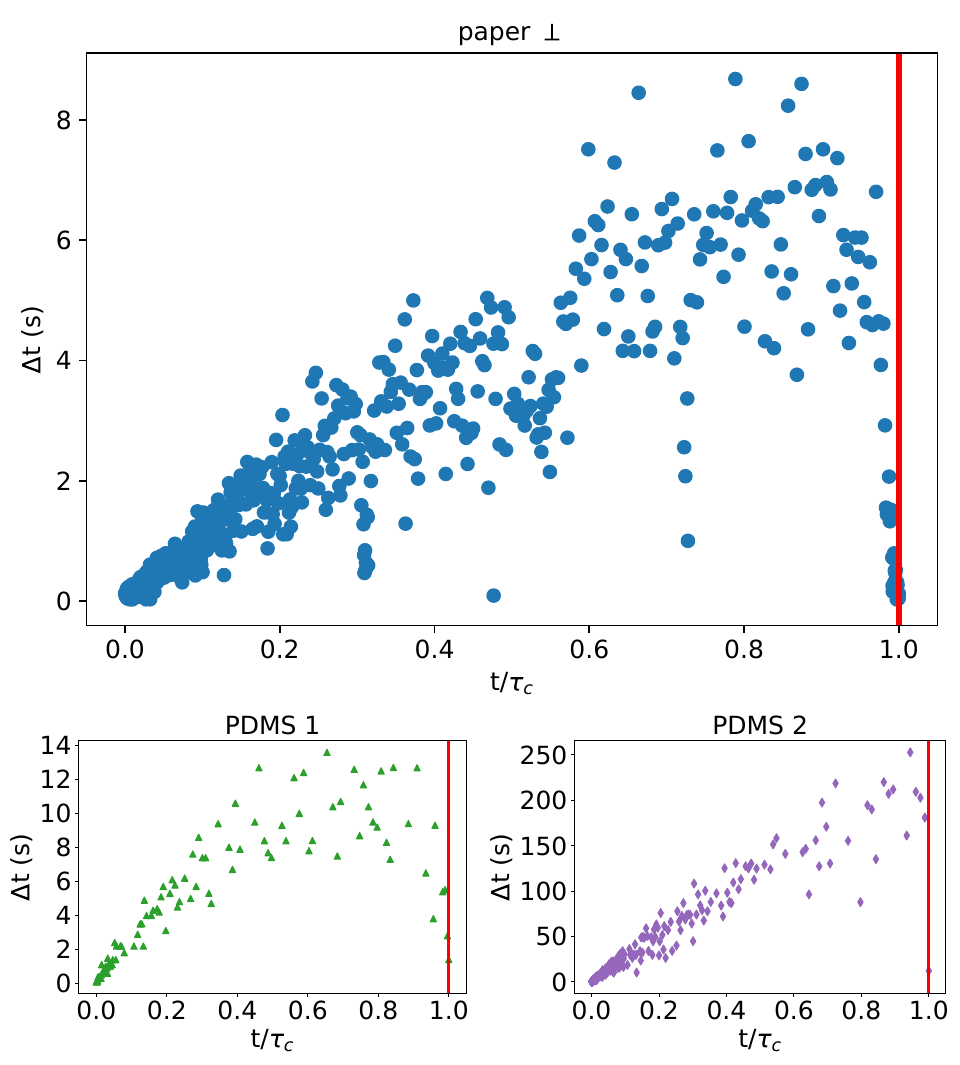}
    \caption{Duration of successive relaxations $\Delta t$ as a function of the normalized time $t/\tau_c$ for paper $\perp$ (top), PDMS~1 (bottom left), and PDMS~2 (bottom right). All samples show a near-linear increase in $\Delta t$ for at least \SI{50}{\percent} of the lifetime $\tau_c$. Paper and PDMS~1 also show a clear decrease before failure.}
    \label{fig:dt evolution}
\end{figure}

As shown on Fig.~\ref{fig:subcritical process}, the stress amplitude $\Delta \sigma$ remains approximately constant during creep, indicating no significant change in the paper’s Young’s modulus. We observe the same stability for the two paper configurations (// and $\perp$) and for the two PDMS samples.
In contrast to the constant stress amplitude $\Delta \sigma$, the relaxation duration $\Delta t$ evolves significantly during creep, indicating that the stress relaxation dynamics depend on the material's age - that is, the time spent creeping under constant stress.
For all samples, $\Delta t$ initially increases progressively over a duration corresponding to at least \SI{50}{\percent} of the total lifetime $\tau_c$.
For paper $\perp$, the increase lasts for $\sim$\qty{80}{\percent} of the sample's lifetime, after which $\Delta t$ drops sharply (see Fig.~\ref{fig:dt evolution}, top). The same trend is observed in the second configuration (paper //).
For the first PDMS sample (\textit{GTeek}), $\Delta t$ increases until about \qty{50}{\percent} of the lifetime and begins to decrease at about \qty{90}{\percent} (see Fig.~\ref{fig:dt evolution}, bottom left). In contrast, for the second PDMS (\textit{Goodfellow}), $\Delta t$ continues to increase until failure (see Fig.~\ref{fig:dt evolution}, bottom right). The absence of a $\Delta t$ decrease for this PDMS may reflect a genuine physical effect or result from the limited temporal resolution of our measurement system, which might miss a rapid acceleration phase at the end of creep.

\begin{figure}[ht!]
    \centering
    \includegraphics[width=0.99\linewidth]{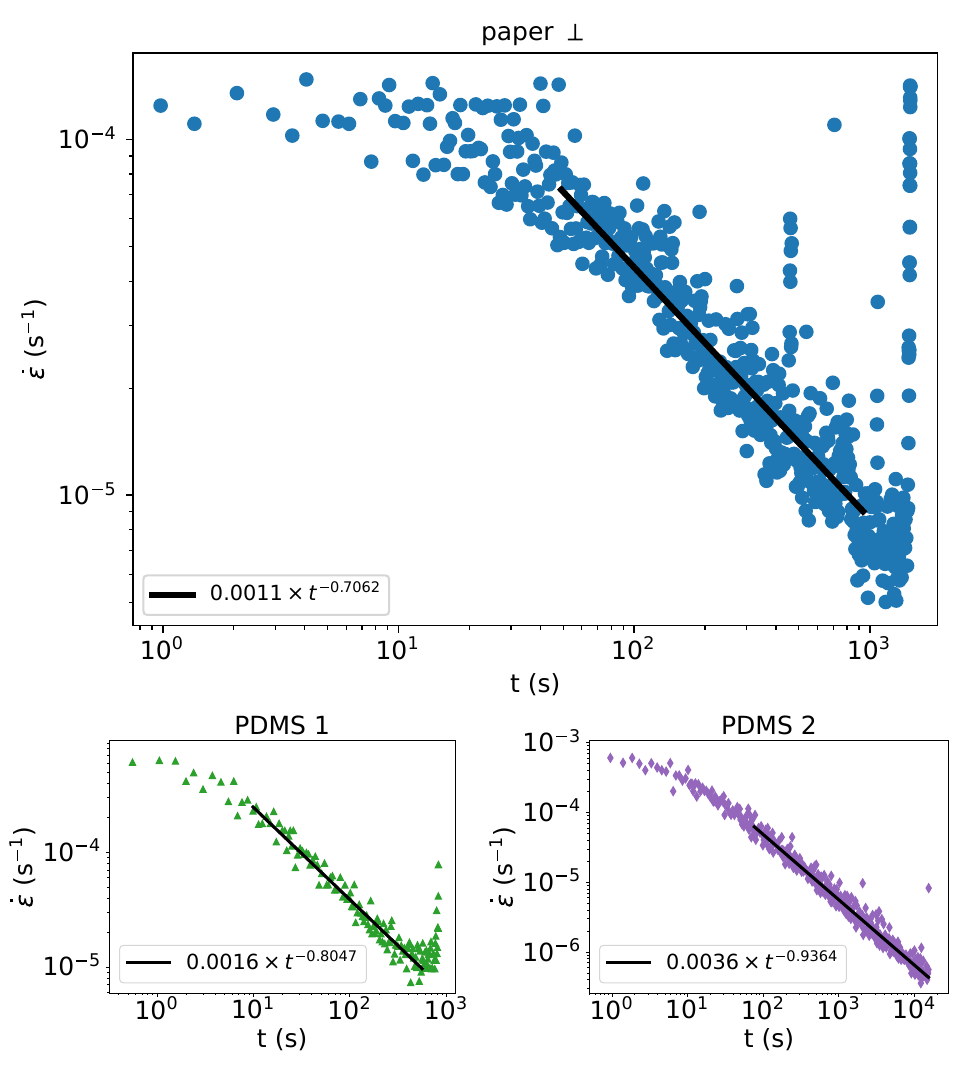}
    \caption{Strain rate $\dot{\varepsilon}$ as function of time for paper (top), PDMS~1 (bottom left), and PDMS~2 (bottom right). For all samples, the primary creep regime follows $\dot{\varepsilon} = \gamma\times t^\alpha$. Fitted values of $\alpha$ and $\gamma$ are shown in each legend.}
    \label{fig:dot_epsilon_materiaux}
\end{figure}

To allow comparison with previous studies on creep-induced strain evolution, we define an effective strain rate $\dot\varepsilon$ from the measurement of $\Delta t$:
\begin{equation}
    \dot\varepsilon = \frac{\varepsilon}{\Delta t},
\end{equation}
where $\varepsilon$ is the strain step after each relaxation ($\varepsilon = \Delta u/L$ with $L$ the sample length).
As shown in Fig.~\ref{fig:dot_epsilon_materiaux}, the initial strain rate decay in paper $\perp$ and the two PDMS samples follows a power law:
\begin{equation}
    \dot\varepsilon = \gamma t^{-\alpha},
    \label{eq:powerlaw}
\end{equation}
with $\gamma$ and $\alpha$ constant. This power law behavior is characteristic of the primary creep regime, observed across several class of materials (metals~\cite{Wyatt1951, Cottrell1952}, composites~\cite{Nechad2005}, paper~\cite{Rosti2010, Koivisto2016}, concrete~\cite{Makinen2023}, protein gels~\cite{Leocmach2014,Bauland2023}, colloidal gels~\cite{Aime2018, Cho2022}, hydrogels~\cite{Karobi2016, Pommella2020}), and numerical simulations~\cite{Hidalgo2002, Jagla2011, Politi2002, Saichev2005, Fusco2013, Roy2022, Weiss2023, Moorcroft2013, Miguel2002, Merabia2016, Lockwood2024}, with exponent values ranging from 0.4 to 1 (see table~\ref{tab:alphas} for a review of $\alpha$ values found in the literature). The coefficient $\gamma$ sets the scale of the strain rate: a higher $\gamma$ corresponds to a faster creep. The exponent $\alpha$ sets the deceleration of the strain rate: a higher $\alpha$ corresponds to a creep that slows down more rapidly with time.

\begin{figure}[ht!]
    \centering
    \includegraphics[width=0.99\linewidth]{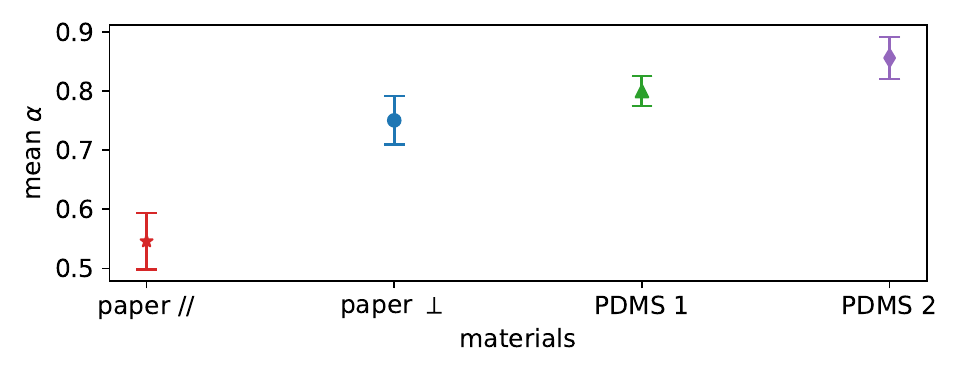}
    \caption{Average exponent $\alpha$ for the four tested materials. Based on 5 (paper //), 9 (paper $\perp$), 9 (PDMS~1), and 6 (PDMS~2) measurements. Errorbars indicate data dispersion.}
    \label{fig:alpha materiaux}
\end{figure}

\begin{figure}[ht!]
    \centering
    \includegraphics[width=0.99\linewidth]{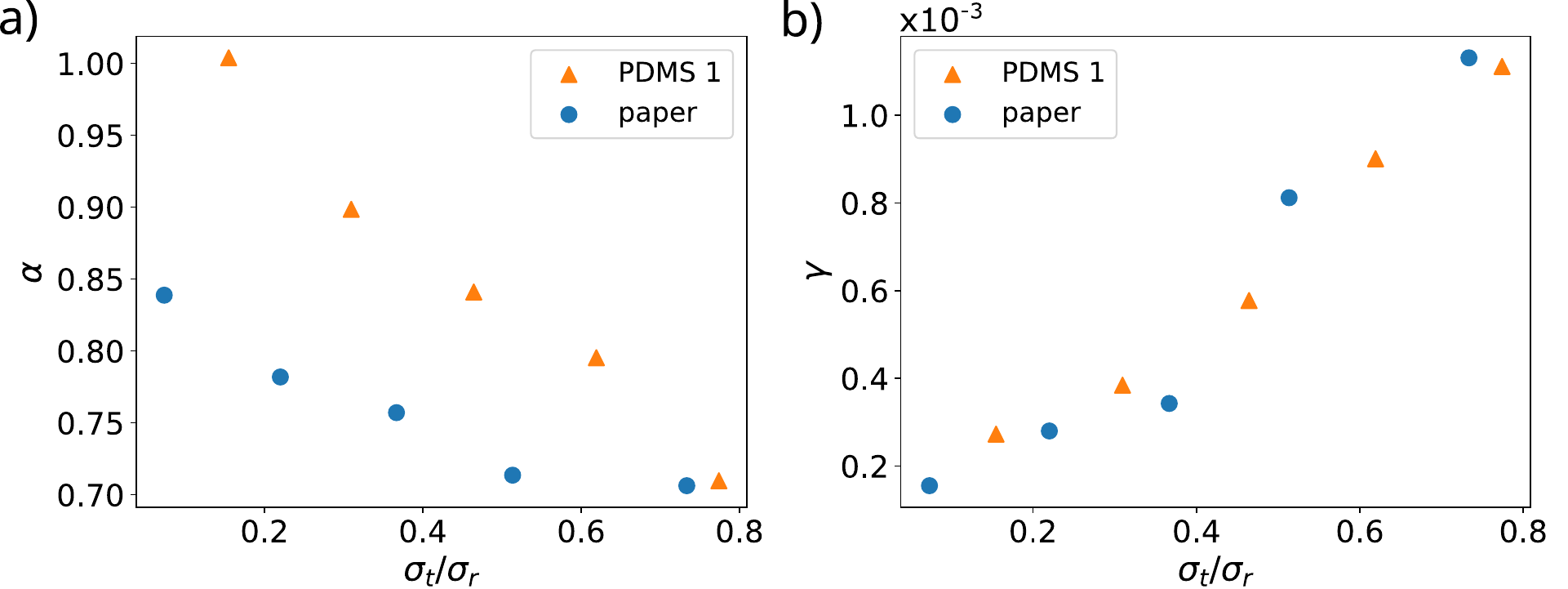}
    \caption{Power law parameters $\alpha$ (a) and $\gamma$ (b) as a function of the normalized stress $\sigma_t/\sigma_r$, for paper $\perp$ and PDMS~1.}
    \label{fig:alpha_et_gamma_exp}
\end{figure}

Across our materials, the averaged exponent $\alpha$ ranges from 0.5 to 0.95 (see Fig.~\ref{fig:alpha materiaux}). Notably, although variation with stress orientation is generally unexpected,  $\alpha$ varies significantly in paper with stress direction relative to the fibers. Fig.~\ref{fig:alpha_et_gamma_exp} shows the evolution of the exponent $\alpha$ and the coefficient $\gamma$ for paper $\perp$ and PDMS~1 as functions of the normalized target stress $\sigma_{t}/\sigma_{r}$. As $\sigma_{t}$ approaches the ultimate tensile stress $\sigma_r$, $\alpha$ decreases while $\gamma$ increases. A similar decrease of $\alpha$ with increasing load has been experimentally observed in uniaxial compression creep of concrete samples~\cite{Makinen2023}. The increase of $\gamma$ indicates that the primary creep accelerates when the target stress is closer to the rupture stress. On the contrary, $\gamma$ is expected to approach zero in the limit of vanishing applied stress.

\begin{table}[ht!]
    \centering
    \begin{tabular}{|c|c|c|}
        \hline
        citation & materials & value \\
        \hhline{|=|=|=|}
        \cite{Andrade1910} & lead & 2/3\\ \hline
        \cite{Wyatt1951} & copper & 2/3\\ \hline
        \cite{Nechad2005} & \makecell{polyglass/polyester \\ composite materials} & $\sim 1$\\ \hline
        \cite{Rosti2010} & paper & 0.75\\ \hline
        \cite{Koivisto2016} & paper & 2/3\\ \hline
        \cite{Makinen2023} & concrete & 0.5 $\leq$ $\alpha$ $\leq$ 0.95 \\ \hline
        \cite{Leocmach2014} & casein gels & 0.85\\ \hline
        \cite{Bauland2023} & casein gels & 0.7\\ \hline
        \cite{Karobi2016} & polyampholyte gel & 0.8 $\leq$ $\alpha$ $\leq$ 1\\ \hline
        \cite{Pommella2020} & biopolymer gel & 0.83\\ \hline
        \cite{Aime2018} & colloidal gel & 0.43\\ \hline
        \cite{Cho2022} & colloidal gel & 0.4 $\leq$ $\alpha$ $\leq$ 0.9\\ \hline
        \cite{Jagla2011} & DFBM with viscoplastic fibers & 1.3 or 1.5\\ \hline
        \cite{Weiss2023} & DFBM with brittle fibers & 2/3 $\leq$ $\alpha$ $\leq$ 1\\ \hline
        \cite{Miguel2002} & Rolie-Poly model & 0.69\\ \hline
        \cite{Lockwood2024} & Soft Glassy Rheology model & 0.3\\
        \hline
    \end{tabular}
    \caption{Values of exponents $\alpha$ found in other studies.}
    \label{tab:alphas}
\end{table}

Those observations can be interpreted in the framework of exhaustion models, where the material is considered as a collection of entities (fibers, sites, cross-links, dislocations, etc.) with disperse mechanical properties, that can break or move. When the material is submitted to an external load, the weakest entities will fail first. Therefore, the creep will initially be relatively fast. However, as time passes, less and less weak entities remain, which induces a decrease of the failure rate. Therefore, the creep slows down until only very strong entities remain. When those entities start to fail, the dynamics starts to accelerate again, as each failure corresponds to a catastrophic loss for the material integrity. It has been recently proposed that a thermally activated DFBM with equal load sharing can explain the load-dependence of the $\alpha$ coefficient~\cite{Weiss2023}. The main idea is that, the load redistribution helps to maintain the creep dynamics: when one fiber fails, the load on all the remaining fibers increases. This is completely consistent with our experimental observations: a higher load $\sigma_t$ induces a faster creep (higher $\gamma$) and a strain rate which slows down less rapidly (lower $\alpha$). The same model~\cite{Weiss2023} also predicts that $\alpha$ should decrease with the temperature or the material disorder, which we have tested numerically.

\section{Numerical simulations of Disordered Fiber Bundle Model}

To compare with our experiments, we implemented the same loading procedure - constant stress maintained through a strain-controlled feedback loop - into an equal load sharing DFBM ~\cite{Pradhan2010} with thermal noise. Such models have already been used to study  lifetimes~\cite{Roux2000, Scorretti2001,Politi2002} and creep dynamics~\cite{Weiss2023} of samples under constant load.

The general principle of the model is the following. We consider a bundle of $N_0$ parallel, purely elastic and brittle fibers, with identical Young’s modulus set to $1$, and with equal load sharing. Each fiber $i$ has an individual constant strength $\sigma_r^i$, and breaks when the load $\sigma^i$ applied to it exceed $\sigma_r^i$. Contrary to other models~\cite{Hidalgo2002,Jagla2011}, we do not consider an intrinsic visco-elasticity of the fibers, and assume that the breakage is instantaneous. To model the heterogeneous nature of the material, the strength values are drawn from a statistical distribution with mean value $1$ and variance $Td$. The value $Td$ hence represents the intrinsic disorder existing among the different fibers (a high $Td$ corresponds to a highly disordered material, while a low $Td$ corresponds to a more homogeneous one). We tested two disorder distributions commonly used in material science: Gaussian and Weibullian~\cite{Alava2006}. The fiber bundle is initially loaded with a total stress $\sigma_0$. Since the model assumes an equal load sharing, the initial individual stress on each fiber is simply $\sigma_0^i = \sigma_0/N_0$. Thermal fluctuations are modelled by adding a Gaussian random stress $\delta\sigma^i$, with mean value $0$ and variance $T$, to each fiber's mean stress at each time step~\footnote{In our model, we consider that the temperature is the main source of fluctuations. In some systems stress fluctuations may arise from other sources such as mechanical noise due to breakage of nearby fibers. In this case $T$ would be considered as an effective temperature.}. These fluctuations can induce failure over time, even when the average stress remains below the fiber’s threshold.

The fiber bundle can evolve under either constant stress or constant strain condition. For constant stress, the total stress is fixed $\sigma = \sigma_0$  and the mean stress per intact fiber increases over time as:
\begin{equation}
\sigma^i(t) =\frac{\sigma_0}{N(t)},
\end{equation}
where $N(t)$ is the number of unbroken fibers at time $t$. This rising load accelerates failure, leading to complete rupture at $t = \tau_c$. Under constant strain, each fiber bears a fixed mean stress $\sigma^i = \sigma_0/N_0$ and the total bundle stress decreases over time as:
\begin{equation}
\sigma(t) = \sigma_0 \times \frac{N(t)}{N_0}.
\end{equation}

To simulate constant stress imposed via a strain-controlled feedback loop, as in our experiments, we first set a target stress $\sigma_t$ and run the simulation at constant strain. As the fibers break, the total stress decreases over time. When it drops below an arbitrary threshold (set to \SI{99.75}{\percent} of $\sigma_t$ in our simulations), the stress is instantly raised again to the value $\sigma_t$~\footnote{To simplify, we make the assumption that the time for the stress to rise to $\sigma_t$ is very short compared to the relaxation duration. This hypothesis is well verified in the experimental data, see inset in Fig.~\ref{fig:subcritical process}.}. After this stress increase, the simulation continues to run at constant strain, until the stress reaches the threshold again, and the process repeats. As in the experimental procedure, this leads to a succession of stress relaxations, whose amplitude $\Delta \sigma$ are constant ($\Delta \sigma = 0.0025\times \sigma_t$ here), and whose duration $\Delta t$ is free to evolve over time. Initially, weak fibers break rapidly, yielding fast stress relaxations and small $\Delta t$. As stronger fibers remain, ruptures are less frequent and $\Delta t$ increases. At time  $\tau_\mathrm{min}$, the failure rate reaches a minimum and $\Delta t$ begins to decrease. This marks the point where each rupture significantly increases the total stress, accelerating failure and eventually leading to the complete bundle breakdown.

\begin{figure}[ht!]
    \centering
    \includegraphics[width=0.99\linewidth]{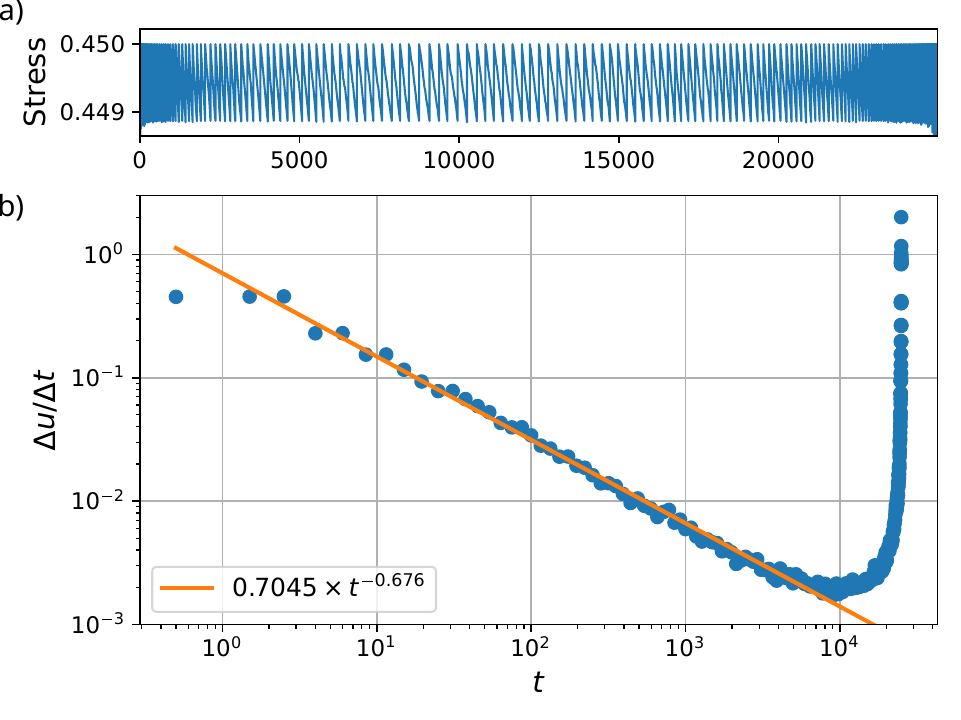}
    \caption{Example of a simulation with Gaussian distribution of strength, $N_0=10^5$, $Td = 0.05$, $T=0.005$, and $\sigma_t = 0.45$. a) Stress as a function of time. b) $\Delta u/\Delta t\propto \dot{\varepsilon}$ as a function of time, with the primary creep regime fitted by the power law indicated in the legend.}
    \label{fig:dot_epsilon_simu}
\end{figure}

An example of a simulation output is shown in Fig.~\ref{fig:dot_epsilon_simu}. The top plot (Fig.~\ref{fig:dot_epsilon_simu}a) shows the time evolution of the stress signal: as expected we observe a succession of relaxations with fixed amplitude $\Delta \sigma$ and variable duration $\Delta t$. Qualitatively, we retrieve the fact that $\Delta t$ first increases with time, and then decreases before the material failure. The bottom plot (Fig.~\ref{fig:dot_epsilon_simu}b) shows, for the same simulation, the time evolution of $\Delta u/\Delta t = \frac{N_0\Delta \sigma}{N(t)\times\Delta t(t)}$ which is proportional to the strain rate $\dot{\varepsilon}$. As in the experiments (see Fig.~\ref{fig:dot_epsilon_materiaux}), the primary creep regime follows a power law $\Delta u/\Delta t = \gamma \times t^{-\alpha}$. Here however, the inflexion point $\tau_\mathrm{min}$ at which the strain rate starts to increase occurs at roughly \SI{50}{\percent} of the total lifetime $\tau_c$.

To understand how this primary creep regime depends on target stress, disorder, and the temperature, we measure the power law parameters $\alpha$ and $\gamma$ for $\sigma_t \in [0.25;0.65]$, $T_d \in \{0.025;0.05;0.1\}$, and $T \in \{0.0025;0.005;0.01\}$. Each set of parameters is run 10 times to increase statistics. For computational efficiency, those simulations were stopped after the first $10^5$ time steps (unless the final failure occurred before $t=10^5$). This duration is long enough to correctly fit the power-law $\Delta u/\Delta t = \gamma \times t^{-\alpha}$ in most cases. To estimate the rupture stress of the bundle $\sigma_r$, we systematically search for the value of target stress $\sigma_t$ at which a complete breakage of the bundle appears almost immediately after starting the simulation~\footnote{Numerically, we consider that a breakage is almost immediate if it occurs in less than 50 time steps.}.

\begin{figure}[ht!]
    \centering
    \includegraphics[width=0.99\linewidth]{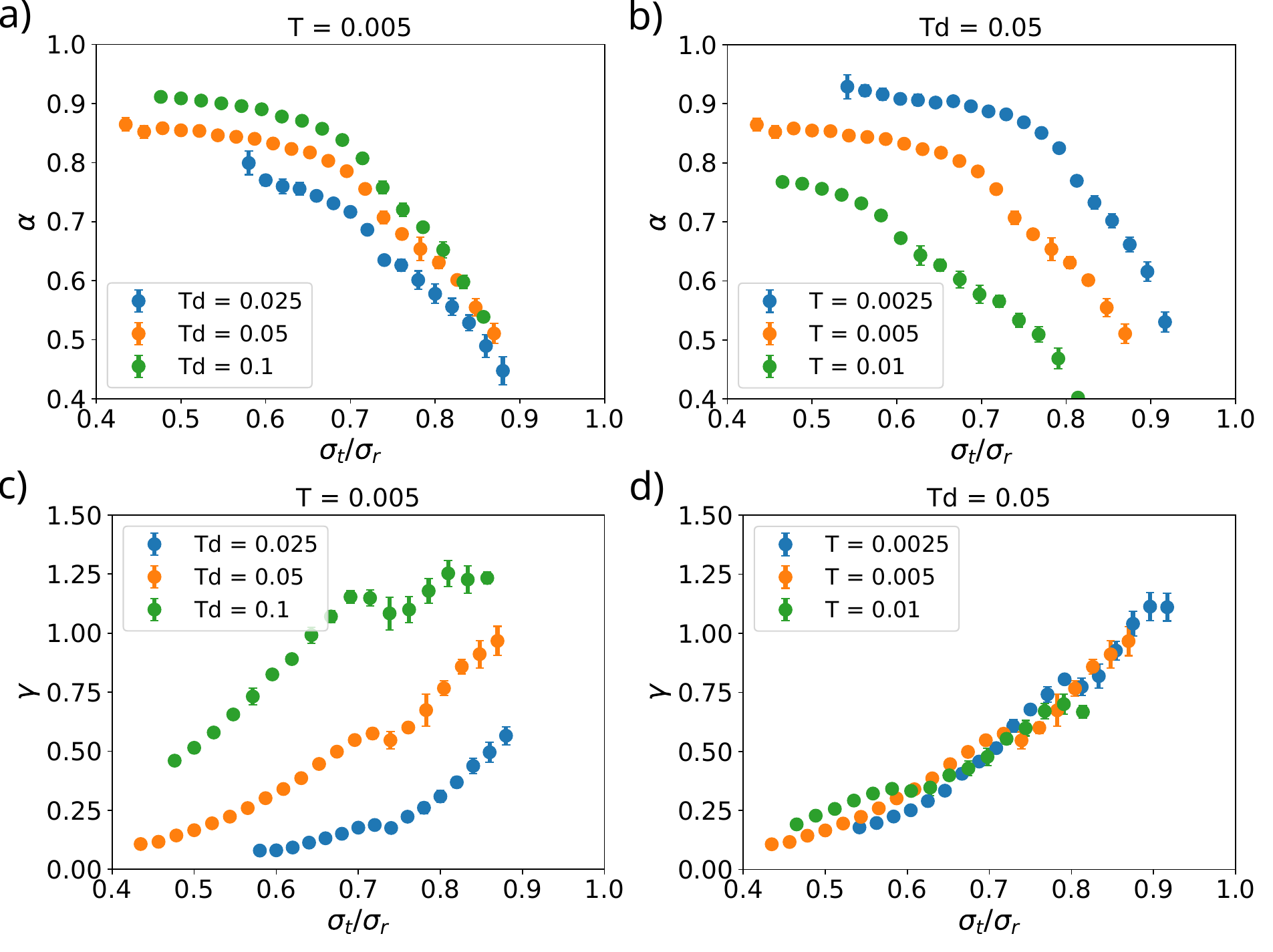}
    \caption{Power-law parameters $\alpha$ (a, b) and $\gamma$ (c, d) as functions of the normalized stress $\sigma_t$/$\sigma_r$, for Gaussian strength distribution, plotted for different disorder $Td$ (a, c) and fluctuation intensity $T$ (b, d). Simulations used $N_0 = 10^5$, with 10 repetition per parameter set. Errorbars indicate standard deviations (sometimes smaller than the symbols).}
    \label{fig:alpha_gamma_simu}
\end{figure}

Results for Gaussian distributions of strengths are shown in Fig.~\ref{fig:alpha_gamma_simu}. As in the experiments (see Fig.~\ref{fig:alpha_et_gamma_exp}), $\alpha$ decreases and $\gamma$ increases as the target stress $\sigma_t$ approaches the rupture stress $\sigma_r$. Notably, $\alpha$ seems to saturate at low $\sigma_t$, which was not observed in the experiment, but may explain why some studies report $\alpha$ as load-independent. The exponent $\alpha$ also increases slightly with disorder $T_d$ (Fig.~\ref{fig:alpha_gamma_simu}a) which is consistent with previous simulations~\cite{Weiss2023}, and may account for the variations observed across materials, or orientations (paper $\perp$ vs paper //). Additionally, $\alpha$ decreases as temperature $T$ increases (Fig.~\ref{fig:alpha_gamma_simu}b), aligning with experiments done on copper~\cite{Wyatt1951} and previous DFBM simulations~\cite{Weiss2023}, but contradicting models for soft gels~\cite{Lockwood2024}, where $\alpha$ increases with temperature. Finally, the coefficient $\gamma$ increases with disorder $T_d$ (Fig.~\ref{fig:alpha_gamma_simu}c) indicating faster stress relaxations in more heterogeneous materials, and shows only a slight change with temperature $T$ (Fig.~\ref{fig:alpha_gamma_simu}d).

\begin{figure}[ht!]
    \centering
    \includegraphics[width=0.99\linewidth]{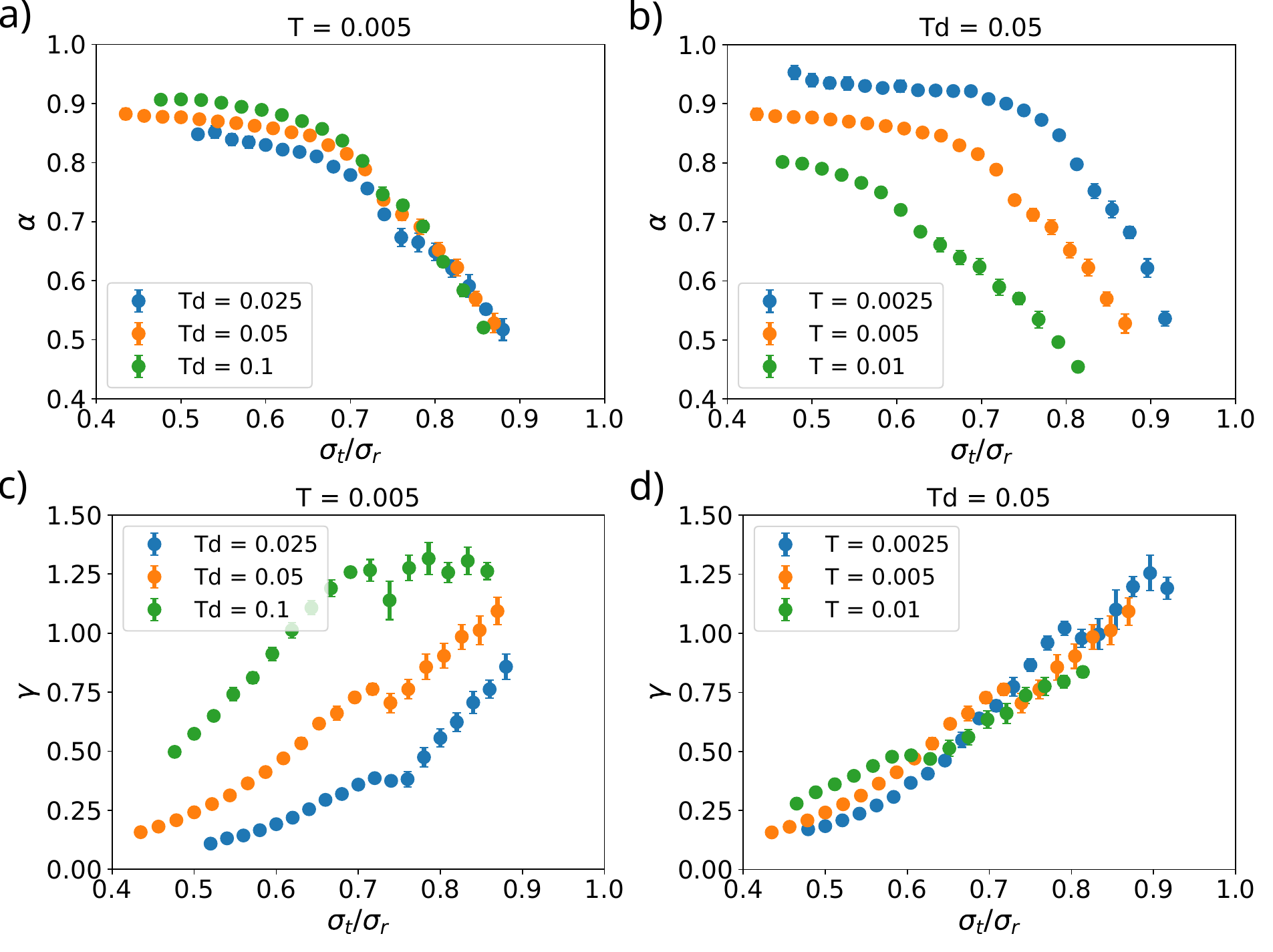}
    \caption{Power-law parameters $\alpha$ (a, b) and $\gamma$ (c, d) as functions of the normalized stress $\sigma_t$/$\sigma_r$, for Weibullian strength distribution, plotted for different disorder $Td$ (a, c) and fluctuation intensity $T$ (b, d). Simulations used $N_0 = 10^5$, with 10 repetition per parameter set. Errorbars indicate standard deviations (sometimes smaller than the symbols).}
    \label{fig:alpha_gamma_simu_weibull}
\end{figure}

Results for Weibullian distributions of strengths are shown in Fig.~\ref{fig:alpha_gamma_simu_weibull}. The scale $\lambda$ and shape $k$ parameters of the distributions were chosen so that they have a mean value $1$ and a variance $Td$ (for example, $Td = 0.05$ corresponds to $\lambda \approx 1.0875$ and $k \approx 5.1334$). The results are very similar to the ones obtained with a Gaussian distribution (see Fig.~\ref{fig:alpha_gamma_simu}), and all our previous observations remain valid. Therefore, we do not believe that the shape of the distributions plays a critical on the dependence of $\alpha$ and $\gamma$ with $\sigma_t$, $T$ or $Td$.

\begin{figure}[ht!]
    \centering
    \includegraphics[width=0.99\linewidth]{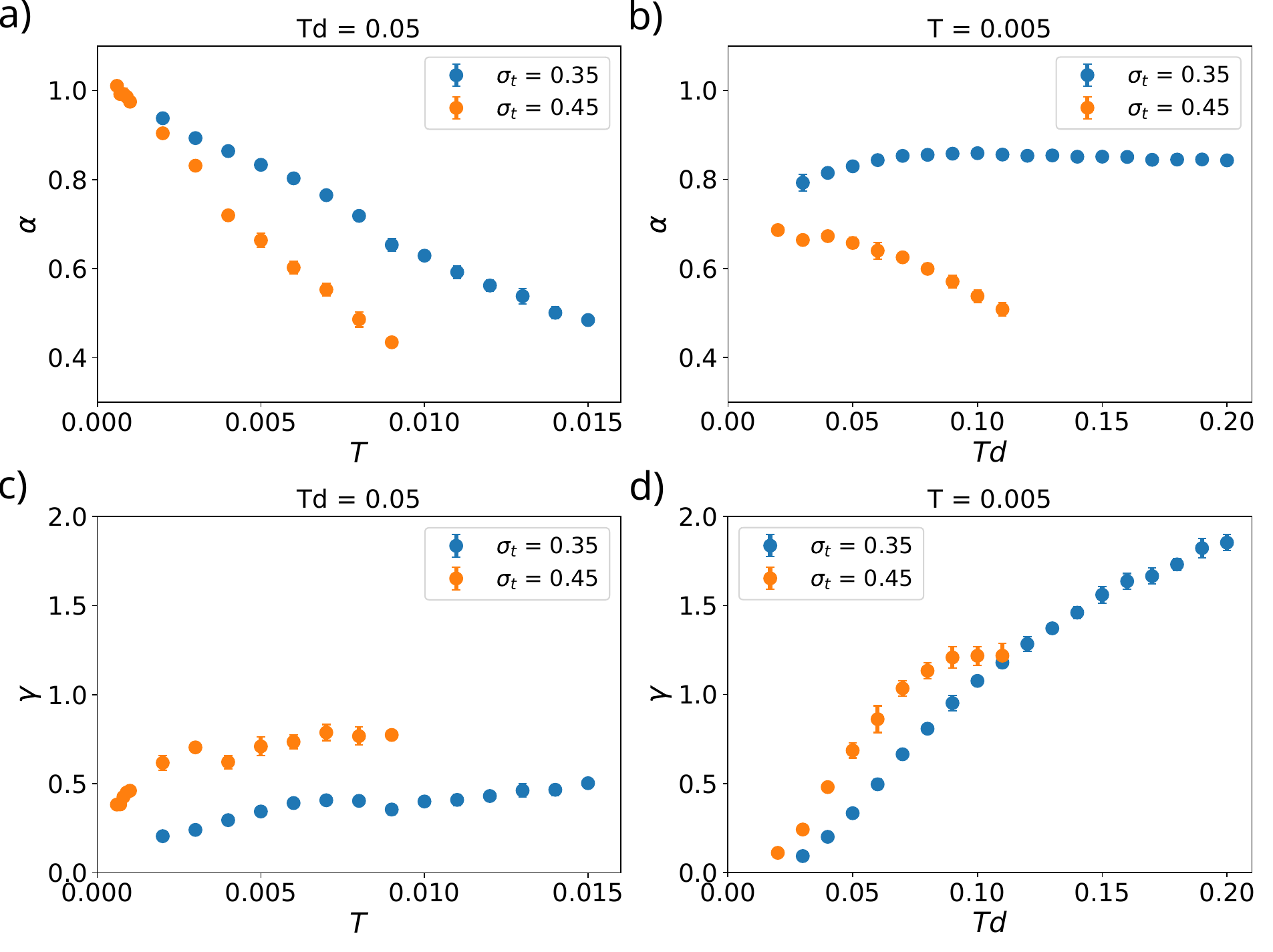}
    \caption{Power-law parameters $\alpha$ (a, b) and $\gamma$ (c, d) as functions of the fluctuation intensity $T$ (a, c) and disorder $Td$ (b, d), for Gaussian strengths distributions. Simulations used $N_0 = 10^5$, with 10 repetition per parameter set. Errorbars indicate standard deviations (sometimes smaller than the symbols).}
    \label{fig:alpha_gamma_simu_T_Td}
\end{figure}

To go further, we also measure $\alpha$ and $\gamma$ as a function of $T$ or $Td$ keeping the other parameters fixed ($\sigma_t \in \{0.35,0.45\}$ and $Td=0.05$ when $T$ is varied, $\sigma_t \in \{0.35;0.45\}$ and $T=0.005$ when $Td$ is varied). The results are shown in Fig.~\ref{fig:alpha_gamma_simu_T_Td}. The general trend corresponds qualitatively to what is expected for a DFBM~\cite{Weiss2023}: as it is the case for the stress $\sigma_t$, increasing the temperature $T$ helps to sustain the creep dynamics, therefore the scale of the strain rate $\gamma$ increases while the exponent $\alpha$ decreases. However, we observe several differences between the effect of the stress, the temperature and the disorder. As seen in Fig.~\ref{fig:alpha_gamma_simu_T_Td}a the temperature has a large effect on the exponent $\alpha$ (which varies from 0.4 to 1), but it seems that the values does not saturate at low $T$. In comparison, the disorder $Td$ seems to have a rather small effect on $\alpha$ (Fig.~\ref{fig:alpha_gamma_simu_T_Td}b). For the lower stress ($\sigma_t = 0.35$), the trend is non-monotonic: $\alpha$ firt slightly increases with $Td$ and then slightly decreases. For the higher stress ($\sigma_t=0.45$), we only observe a small decrease of $\alpha$~\footnote{Note that this is not in contradiction with what is shown in Fig.~\ref{fig:alpha_gamma_simu}a, because $\sigma_r$ also depends on $T$ and $Td$.}. For the coefficient $\gamma$, the relative importance of $T$ and $Td$ seems reversed: $T$ has a rather small effect on $\gamma$ (Fig.~\ref{fig:alpha_gamma_simu_T_Td}c), while $Td$ has a larger effect on it (Fig.~\ref{fig:alpha_gamma_simu_T_Td}d).

In terms of primary creep dynamics, the stress $\sigma_t$, the temperature $T$, and the disorder $Td$ do not have the same effect. Both the stress and the disorder have a large impact on the scale of the strain rate $\gamma$: a higher $\sigma_t$, or a higher $Td$ in the material will result in a faster creep. This comes from the fact that a higher stress or a broader distribution of fibers strengths will result in a lower energy barrier for the weak fibers to break. In contrast, the scale of the strain rate $\gamma$ is only weakly dependent on the temperature. This means that changing $T$ has little effect on the speed at which the material will initially creep. However, $T$ strongly affects the exponent $\alpha$ controlling the slowing down of the creep dynamics. This can be explained by the fact that it becomes more probable that strong fibers can break before weaker ones at higher temperatures. Indeed, in the limit of very high $T$, each fiber has approximately the same probability to break at each time step, regardless of its strength. Therefore, it becomes easier for the creep to sustain itself at the same speed and the deceleration of the dynamics, which comes from the fact that weak fibers break first, is less pronounced. Finally, we note that for a given set of $T$ and $Td$ the power-law exponent $\alpha$ saturates for stresses below a given threshold (see Fig.~\ref{fig:alpha_gamma_simu}(a,b) and Fig.~\ref{fig:alpha_gamma_simu_weibull}(a,b)). Moreover, the stress threshold, and more importantly, the value at which $\alpha$ saturates seem to depend on $T$ and $Td$~\footnote{Their full dependency is beyond the scope of this article.}, which may explain why other experimental studies have found different load-independent $\alpha$ values for different materials. If the range of applied stresses is too far away from the rupture stress $\sigma_r$, it is possible to find values of $\alpha$ which do not evolve significantly with the stress.

\section*{Conclusions}

In conclusion, we investigated subcritical rupture of heterogenous materials under constant load, confirming a primary creep regime where the strain rate decreases as a power-law $\dot{\varepsilon} = \gamma t^{-\alpha}$, before accelerating towards failure after time $\tau_\mathrm{min}$. This behavior, observed for all tested materials (paper samples in two orthogonal load directions, and PDMS samples from two different manufacturers), is captured by a thermally activated DFBM. Except for one type of samples, the inflection point $\tau_\mathrm{min}$ occurs between \SI{50}{\percent} and \SI{80}{\percent} of the lifetime, and can hence be seen as a precursor of the final rupture. Both parameters $\alpha$ and $\gamma$ depend not only on the material, but also on the applied stress $\sigma_t$ and the loading direction in the case of an anisotropic material. Experimentally, $\alpha$ decreases and $\gamma$ increases with increasing $\sigma_t$. These observations are confirmed numerically using the DFBM, where the effects of disorder $T_d$ and temperature $T$ on $\alpha$ and $\gamma$ were also studied. Notably, for a given set of parameters, we find the existence of a threshold stress value below which the exponent $\alpha$ becomes nearly constant, which may explain previous reports of a stress-independent exponents.

\section*{Data availability}
The data (experimental data, numerical data, simulations and analysis Python codes) that support the findings of this article are openly available~\cite{DataZenodo}.

\begin{acknowledgments}
This work was supported by the French National Research Agency grant n$^\circ$ANR-22-CE30-0019. The authors would like to thank Matthieu Mercury for his help with repairing and improving the tensile apparatus set up.
\end{acknowledgments}

\bibliography{biblio}

\end{document}